\documentclass[preprint,chaos,amsmath,amsthm,amssymb,showpacs,floatfix]{revtex4-1}
\usepackage{graphicx}
\usepackage{dcolumn}
\usepackage{bm}
\newcommand{\be}{\begin{equation}}
\newcommand{\ee}{\end{equation}}

\newtheorem{theorem}{Theorem}

\newtheorem{conj}[theorem]{Conjecture}

\def\proof{{\bf Proof. }}

\begin{document}
\title{ On Fractional Eulerian Numbers and Equivalence of  Maps with 
Long Term Power-Law Memory (Integral Volterra Equations of the Second Kind) to
Gr$\ddot{u}$nvald-Letnikov Fractional Difference (Differential) Equations 
}

\author{M. Edelman}

\affiliation{Department of Physics, Stern College at Yeshiva University, 245
  Lexington Ave, New York, NY 10016, USA 
\\ Courant Institute of
Mathematical Sciences, New York University, 251 Mercer St., New York, NY
10012, USA
}

\date{\today}

\begin{abstract}
In this paper we consider a simple general form of a deterministic 
system with power-law memory whose state can be described by one variable 
and evolution by a generating function. 
A new value of the system's variable is 
a total (a convolution) of the generating functions of all previous values
of the variable with weights, which are powers of the time passed. 
In discrete cases these systems can be described by difference equations
in which a fractional difference on the left hand side is equal to a
total (also a convolution) 
of the generating functions of all previous values of the system's 
variable with fractional Eulerian number weights
on the right hand side. In the continuous limit the considered systems
can be described by Gr$\ddot{u}$nvald-Letnikov fractional differential 
equations, which are equivalent to the Volterra integral equations of the 
second kind. New properties of fractional Eulerian numbers 
and possible applications of the results are discussed.
\end{abstract}
\maketitle



{\bf 
}
\section{Introduction}
\label{int}

In paper \cite{DNC} we introduced $\alpha$-families of maps ($\alpha$FM) 
which correspond to a general form of fractional differential 
equations of systems experiencing periodic kicks
\be
\frac{d^{\alpha}x}{dt^{\alpha}}+\tilde{G}_K(x(t- \Delta T)) \sum^{\infty}_{k=-\infty} \delta \Bigl(\frac{t}{T}-(k+\varepsilon)
\Bigr)=0,   
\label{UM1D2Ddif}
\ee
where $\tilde{G}_K(x)$ is an arbitrary non-linear function, K is a parameter, 
$\varepsilon > \Delta > 0$,  $\alpha \in \mathbb{R}$, $\alpha>0$, in
the limit $\varepsilon  \rightarrow 0$, with the initial conditions
corresponding to the type of the fractional derivative used. We
investigated their general properties in \cite{DNC} and the following 
articles \cite{Chaos,Chaos2014,DNC2014,ICFDA2014}. 
These maps are maps with power-law memory in which the new value 
of the variable $x_{n+1}$ depends on all previous values $x_{k}$ 
$(0 \le k \le n)$ of the same variable with weights proportional 
to the time passed $(n+1-k)$ to the power $(\alpha-1)$.
For example, in the case of the Caputo fractional derivatives
Eq.~(\ref{UM1D2Ddif}) leads to (for $T=1$)
\be
x_{n+1}= \sum^{N-1}_{k=0}\frac{x^{(k)}_0}{k!}(n+1)^{k} 
-\frac{1}{\Gamma(\alpha)}\sum^{n}_{k=0} \tilde{G}_K(x_k) (n-k+1)^{\alpha-1},
\label{FrCMapx}
\ee 
where $x^{(k)}(t)=D^k_tx(t)$, $x^{(k)}_0=x^{(k)}(0)$, $0 \le
 N-1 < \alpha \le N$, $\alpha \in \mathbb{R}$, $N \in \mathbb{N}$.

Historically, the first maps with memory 
were considered as models for  non-Markovian processes in general 
\cite{MM3,MM4} and, with regards  to thermodynamic theory of systems 
with memory \cite{MM5}, as analogues of the integro-differential
equations of non-equilibrium statistical physics \cite{MM1,MM2,MM6}.
The general form of the investigated maps was
\begin{equation}
x_{n+1}=\sum^{n}_{k=m}V(n,k)G(x_k),
\label{LTM}
\end{equation} 
where $V(n,k)$ characterizes memory effects.  Maps Eq.~(\ref{LTM}) 
with $m=0$
are called maps with long term memory. 
Maps  in which the number of
terms in the sum in  Eq.~(\ref{LTM}) is bounded ($m=n-M+1$) 
are called maps with short term memory or  M-step memory maps. 

In this paper we consider long term memory maps 
with power-law memory in the form 
\begin{equation}
x_{n}=\sum^{n-1}_{k=0}(n-k)^{\alpha-1} G_K(x_k,h),
\label{LTMPL}
\end{equation}
where $K$ is a parameter and $h$ is a constant time step between 
$t_n$ and $t_{n+1}$. These maps differ from the maps
Eq.~(\ref{FrCMapx}) by the sum of power functions depending 
on the initial conditions of Eq.~(\ref{UM1D2Ddif}). 
They coincide in the case of the zero initial
conditions, $h=1$, and $G_K(x_k) = -\tilde{G}_K(x_k)/\Gamma(\alpha)$.

Interest in power-law memory maps is stimulated by the
recent discovery of the large number of systems (mostly biological),  
not necessarily described by the fractional differential equations, 
with power-law memory. In the study of human memory, 
the accuracy on a memory tasks, decays 
as a power law,  $\sim t^{-\beta}$, with $0<\beta<1$ 
\cite{Kahana,Rubin,Wixted1,Wixted2,Adaptation1}. 
In the study of human learning, the reduction in reaction times that comes with
practice is a power function of the number of training trials \cite{Anderson}.
Power-law adaptation has been used 
to describe the dynamics of biological systems in 
\cite{Adaptation1,Adaptation3,Adaptation4,Adaptation2,Adaptation5,Adaptation6}.
As it has been shown recently, even processing of external stimuli 
by individual neurons can be described by fractional 
differentiation \cite{Neuron3,Neuron4}.
Most of human organ tissues demonstrate viscoelastic properties 
\cite{TissueNerv,TissueBrain2,TissueBrain1,Coussot,TissueLiver2,TissueLiver1,TissueSpleen,TissueProstate1,TissueProstate2,TissueArteries1,TissueArteries2,TissueMuscle}. 
This leads to their description by  fractional differential equations
with time fractional derivatives \cite{Podlubny,MainardiBook2010,CM1971,Visc4,BT1983a,BT1983b,MG2007,Visc5,Visc6} which implies the power-law memory.
In most of the biological systems with the power-law  behavior ($\sim t^\beta$)
the power $\beta$ is between $-1$ and and $1$, which leads to  
$0<\alpha<2$ in Eq.~(\ref{LTMPL}).

Biological systems are not the only natural systems with 
power-law memory. In the continuous case
these systems  can be described by fractional
differential equations and one may find many examples of such systems in
the recent books on applications of fractional calculus 
\cite{MainardiBook2010,Samko,KilBook,ZasBook2008,FrDyn2011,TarasovBook2011,UchSib,ControlBook2010,ControlBook2011,ControlBook2012,LuoBook2010,Nour,Fields,LM}.
In physics, for example,  common and general examples of systems with 
power-law  memory 
include:   Hamiltonian systems, in which transport can be described by
the fractional Fokker-Plank-Kolmogorov equation 
and memory is the result of stickiness of 
trajectories in time to the islands of regular motion,
\cite{ZasBook2008,ZE1997,ZE2000,ZE2004}; 
dielectric materials, where electromagnetic fields  are described by
equations with time fractional derivatives due to the ’universal’ 
response - the power-law frequency dependence of the dielectric 
susceptibility in a wide range of frequencies 
\cite{TarasovBook2011,TD1,TD2,TD3}; 
materials with rheological properties and viscoelastic materials, in which 
non-integer order differential stress-strain relations give a minimal 
parameter set concise description of polymers and other viscoelastic 
materials with non-Debye relaxation and memory of strain history 
\cite{MainardiBook2010,CM1971,BT1983a,BT1983b,MG2007}. 
It is also interesting that the use fractional calculus (power-law memory) 
in control (fractional order control) makes it possible to improve
performance of traditional controllers \cite{ControlBook2010,ControlBook2012}.

Another motivation for the present paper comes from the first results of the
investigation of fractional (power-law memory, see e.g., 
Eq.~(\ref{FrCMapx}) 
\cite{DNC,Chaos,Chaos2014,DNC2014,ICFDA2014,ME1,ME2,ME3,ME4,ME5})
and fractional difference (asymptotically power-law memory 
\cite{Chaos2014,DNC2014}) maps. 
\begin{figure}[b]
\begin{center}
\includegraphics[scale=.9]{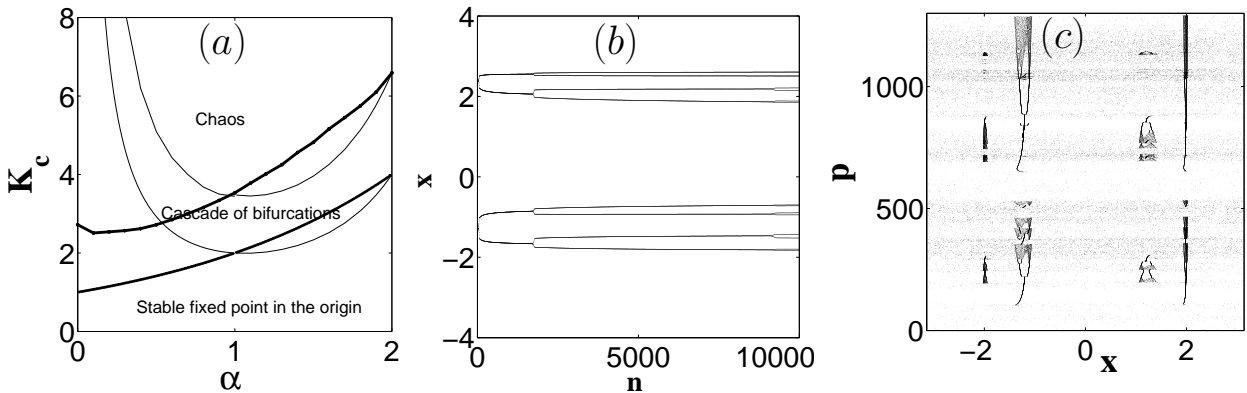}
\end{center}
\caption{Bifurcations and cascade of bifurcations type trajectories in
  fractional/(fractional difference) maps: (a). $\alpha$-$K$ diagrams for
  the Caputo fractional (thin lines) and fractional difference (bold
  lines) Standard Maps (see \cite{Chaos2014}). Memory parameter $\alpha$ 
   corresponds to the  $\alpha$ in Eq.~(\ref{LTMPL}) and $K$ is a
   non-linearity parameter, which in the case  $\alpha=2$ coincides with
   the non-linearity parameter in the regular Standard Map \cite{Chirikov}.
Fixed point in the origin is stable below the lower curves and chaos
exists above the upper curves. Period doubling cascades of bifurcations
occur between the lower and upper curves; (b). A single trajectory (CBTT)
for the  Caputo fractional difference Standard Map with $\alpha=0.1$,
$K=2.4$, and the initial condition $x_0=0.1$; (c).  A single trajectory
(intermittent CBTT) for the Riemann-Liouville fractional Standard Map with 
 $\alpha=1.557$ and $K=4.21$.  
 }
\label{Fig1}      
\end{figure}
It has been shown that fractional and fractional difference maps both 
demonstrate new type of attractors - cascade of bifurcations type 
trajectories (CBTT) (see Fig.~\ref{Fig1}) in which after a small number 
of iterations a trajectory converges to a period one trajectory (fixed
point) which later bifurcates and becomes a $T = 2$ sink and then follows 
the period doubling scenario typical for cascades of bifurcations in 
regular dynamics. The difference is that in regular dynamics a cascade 
of bifurcations is the result of a change in a non-linearity parameter
and in CBTT a cascade of bifurcations occurs on a single attracting
trajectory. CBTT were demonstrated in the examples of harmonic
and quadratic maps with power-law (and falling factorial-law, which is
asymptotically power-law) memory derived from differential
equations with the Riemann-Liouville and Caputo fractional derivatives
(and from Caputo fractional difference equations) with $\alpha \in (0,2)$.
In regular continuous dynamical systems the Poincar$\acute{e}$-Bendixson 
theorem shows that chaos can only arise in systems with more than two 
dimensions. This is a consequence of the fact that phase space trajectories
can't intersect. Dependence of solutions of  fractional
differential equations on the whole history of the 
corresponing system's evolution 
makes intersection of trajectories possible (see Fig.~\ref{SI} and 
one may consider a conjecture that chaos and CBTT are 
possible in fractional systems 
with less than two dimensions. One of the goals of the present paper is to
investigate a possibility of preserving chaotic behavior during a
transition from discrete to continuous fractional systems in less than two
dimensions. 
\begin{figure}[b]
\begin{center}
\includegraphics[scale=.9]{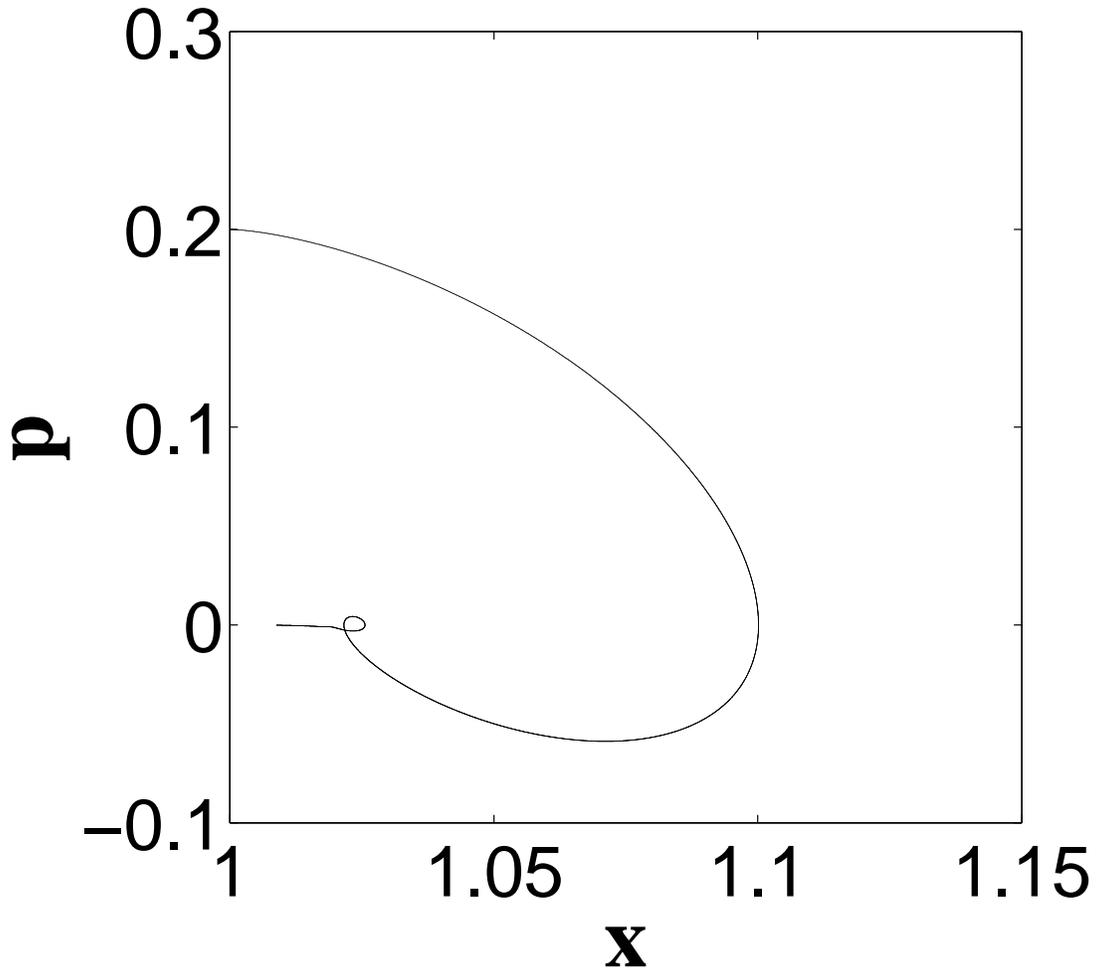}
\end{center}
\caption{A self intersecting phase space trajectory of the 
fractional Caputo Duffing equation 
$_0^CD^{1.5}_t x(t)=x(1-x^2)$, $t\in [0,40]$ 
with the initial conditions $x(0)=1$
and $dx/dt(0)=0.2$. For the definition of the fractional Caputo derivative see 
Eq.~(\ref{Cap}).
} 
\label{SI}      
\end{figure}

There is also a fundamental question of the origin of the Universe and
a related question of the origin of the memory of living species. Were
there seeds of memory present at the origin of the Universe?
Were the fundamental laws of nature memoryless or did they have some form
of memory? One of the approaches is to assume that on the time and length 
scales smaller than Planck time and length the fundamental laws 
should have some memory and a feedback mechanism in order to manage its
evolution. This is a purely philosophical question unless we show 
that the presence of memory may lead to a fundamentally different 
behavior of the Universe on the large scales and compare it with the
observations. This is yet another motivation to investigate the very basic
properties of systems with memory.

In what follows we prove  the equivalence of the map
Eq.~(\ref{LTMPL}) with the non-negative integer power-law memory ($\alpha=m>0$) 
to the m-step memory map in Sec.~ \ref{Integer} and prove a similar theorem
for the maps with $\alpha \in  \mathbb{R}$ in Sec.~ \ref{Real}. 
In Sec.~ \ref{Beh} we consider behavior of the discrete maps with
power-law memory and  transition to the continuous limit as 
$h \rightarrow 0$; in this section we also discuss some properties of
the fractional Eulerian numbers. In Secs.~ \ref{Summary}~and~ \ref{Conclusion}
we summarize our results and discuss their possible applications.

\section{Maps with Non-negative Integer Power-Law Memory}
\label{Integer}

If we assume $\alpha=1$, then the map Eq.~(\ref{LTMPL}) for $n>0$ is equivalent
to 
\begin{equation}
x_1=G_K(x_0,h), \  \  \  x_{n}-x_{n-1}= G_K(x_{n-1},h), \  \    (n>1)
\label{alp1}
\end{equation}
and requires one initial condition $x_0$. Calculation of the second 
backward difference 
from Eq.~(\ref{LTMPL}) for $x_n$ in the case  $\alpha=2$  for $n>0$ yields
\begin{equation}
x_1=G_K(x_0,h), \ \ x_2=2G_K(x_0,h)+G_K(x_1,h), \ \ x_{n}-2x_{n-1}+x_{n-2}= G_K(x_{n-1},h),  \ \    (n>2)
\label{alp2}
\end{equation} 
with the initial condition   $x_0$.
It is easy to see that for   $\alpha=3$ ($n>3$) and $\alpha=4$ ($n>4$)
calculating the third and the fourth backward differences for $x_n$ we obtain
correspondingly 
\begin{eqnarray}
&& x_1=G_K(x_0,h), \  \ x_2=4G_K(x_0,h)+G_K(x_1,h), \  \ 
x_3=9G_K(x_0,h)+4G_K(x_1,h)+G_K(x_2,h),  \nonumber \\
&& x_{n}-3x_{n-1}+3x_{n-2}-x_{n-3}= G_K(x_{n-1},h)+ G_K(x_{n-2},h), \   \ (n>3)
\label{alp3}
\end{eqnarray} 
 and
\begin{eqnarray}
&& x_1=G_K(x_0,h), \ \ 
x_2=8G_K(x_0,h)+G_K(x_1,h), \ \ x_3=27G_K(x_0,h)+8G_K(x_1,h)+G_K(x_2,h),
\nonumber \\
&& x_4=64G_K(x_0,h)+27G_K(x_1,h)+8G_K(x_2,h)+G_K(x_3,h),  \label{alp4} \\
&&x_{n}-4x_{n-1}+6x_{n-2}-4x_{n-3}+x_{n-4}=
G_K(x_{n-1},h)+4G_K(x_{n-2},h)+G_K(x_{n-3},h), \  \  (n>4). \nonumber
\end{eqnarray}
Corresponding summations of 
Eqs.~(\ref{alp1})~(\ref{alp2})~(\ref{alp3})~(\ref{alp4}) 
with weights $(n-k)^{\alpha-1}$ yield Eq.~(\ref{LTMPL}).

Based on Eqs.~(\ref{alp1})-(\ref{alp4}) we may expect the following
theorem:
\begin{theorem}
Any long term memory
map 
\begin{equation}
x_{n}=\sum^{n-1}_{k=0}(n-k)^{m-1} G_K(x_k,h), \ \ (n>0),
\label{IntLTMPL}
\end{equation}
where $m \in  \mathbb{N}$, is equivalent to the $m$-step memory map 
\begin{eqnarray}
&& x_n=\sum^{n-1}_{k=0}(n-k)^{m-1}G_K(x_{k},h),  \ \ \ (0 < n \le m), \nonumber   \\
&&\sum^{m}_{k=0} (-1)^k 
\left( \begin{array}{c}
m \\ k
\end{array} \right)
x_{n-k}=
\delta_{m-1} G_K(x_{n-1},h)+ \sum^{m-2}_{k=0}A(m-1,k)G_K(x_{n-k-1},h), (n > m) 
\label{Mstep}
\end{eqnarray}
\label{The1}
\end{theorem}
In Eq.~(\ref{Mstep}) the alternating sum on the left hand side (LHS) 
is the $m^{th}$ backward
difference for the $x_n$;  $\delta_i$ is the Kronecker delta ($\delta_0=1$;  
$\delta_{i\ne0}=0$); $A(n,k)$ are the Euleruan numbers 
\begin{equation}
A(n,k)=\sum^{k}_{j=0} (-1)^j 
\left( \begin{array}{c}
n+1 \\ j
\end{array} \right)
(k+1-j)^n
\label{EN}
\end{equation}
defined for $k,n \in \mathbb{N}_0$ ($ \mathbb{N}_0:= \mathbb{N} \cup
\{0\}$)
which satisfy the recurrence formula
\begin{equation}
A(n,k)=(k+1)A(n-1,k)+(n-k)A(n-1, k-1).
\label{ENRec}
\end{equation}
\begin{figure}[b]
\begin{center}
\includegraphics[scale=.35]{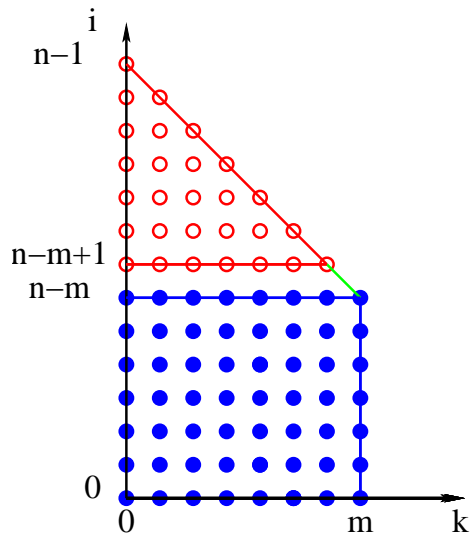}
\end{center}
\caption{The area of summation.
  }
\label{sum}      
\end{figure}
\proof
1. To prove that Eq.~(\ref{IntLTMPL}) leads to Eq.~(\ref{Mstep}) we modify the left side of   Eq.~(\ref{Mstep}) using
Eq.~(\ref{IntLTMPL}):
\begin{equation}
\sum^{m}_{k=0} (-1)^k 
\left( \begin{array}{c}
m \\ k
\end{array} \right)
x_{n-k}=
\sum^{m}_{k=0} (-1)^k 
\left( \begin{array}{c}
m \\ k
\end{array} \right)
\sum^{n-k-1}_{i=0}(n-k-i)^{m-1}G_K(x_{i},h)=S_1+S_2, 
\label{MstepN}
\end{equation}
where $S_1$ and $S_2$ are the sums taken over the points in the upper
triangular and the bottom rectangular areas in Fig.~\ref{sum} correspondingly. 
After changing the order of summation in $S_1$ we have:
\begin{equation}
S_1 = 
\sum^{n-1}_{i=n-m} G_K(x_{i},h)
\sum^{n-1-i}_{k=0}
(-1)^k 
\left( \begin{array}{c}
m \\ k
\end{array} \right)
(n-k-i)^{m-1}. 
\label{S1}
\end{equation}
After introduction $j=n-i-1$ we have
\begin{eqnarray}
&&S_1 = 
\sum^{m-1}_{j=0} G_K(x_{n-j-1},h)
\sum^{j}_{k=0}
(-1)^k 
\left( \begin{array}{c}
m \\ k
\end{array} \right)
(j+1-k)^{m-1}=\sum^{m-1}_{j=0}A(m-1,j)  G_K(x_{n-j-1},h) \nonumber \\  
&&=\delta_{m-1} G_K(x_{n-1},h)+ \sum^{m-2}_{k=0}A(m-1,k)G_K(x_{n-k-1},h).
\label{S1n}
\end{eqnarray}
Here we took into account that according to Eq.~(\ref{Am-1}) below
\begin{equation}
A(m-1,m-1)=\delta_{m-1}.
\label{Ajj}
\end{equation}

For the second sum we have
\begin{equation}
S_2 = 
\sum^{n-m-1}_{i=0} G_K(x_{i},h)
\sum^{m}_{k=0}
(-1)^k 
\left( \begin{array}{c}
m \\ k
\end{array} \right)
(n-k-i)^{m-1}=\sum^{n-m-1}_{i=0} G_K(x_{i},h)S_3(m,n-i), 
\label{S2}
\end{equation}
where  
\begin{equation}
S_3(m,j) = 
\sum^{m}_{k=0}
(-1)^k 
\left( \begin{array}{c}
m \\ k
\end{array} \right)
(j-k)^{m-1}
\label{S3}
\end{equation}
and ($m+1 \le j \le n$).

Let's show that $S_3(m,j) =0$:
\begin{eqnarray}
&&S_3(m,j)= 
\sum^{m}_{k=0}
(-1)^k 
\left( \begin{array}{c}
m \\ k
\end{array} \right)
(j-k)^{m-1}
= 
\sum^{m}_{k=0}
(-1)^k 
\left( \begin{array}{c}
m \\ k
\end{array} \right)
\sum^{m-1}_{i=0}
(-1)^ik^ij^{m-1-i}
\left( \begin{array}{c}
m-1 \\ i
\end{array} \right)
\nonumber \\ 
&&=\sum^{m-1}_{i=0}
(-1)^i
j^{m-1-i}
\left( \begin{array}{c}
m-1 \\ i
\end{array} \right)
S_4(m,i)=0
\label{S31}
\end{eqnarray} 
because
\begin{equation}
S_4(m,i) = 
\sum^{m}_{k=0}
(-1)^k 
\left( \begin{array}{c}
m \\ k
\end{array} \right)
k^i
=
\left\{ 
\begin{array}{ll}
0,  & \mbox{if $0 \le i <m$}, \\ m!(-1)^m, &  \mbox{if $i=m$}. 
\end{array} \right .
\label{S4}
\end{equation}
A simple proof of Eq.~(\ref{S4}) by induction can be found in \cite{Gould} and
a very elegant and short proof  using generating functions can be
found on page 13 of \cite{Riordan}.

For $m>1$
\begin{equation}
A(m-1,m-1) = 
\sum^{m-1}_{k=0}
(-1)^k 
\left( \begin{array}{c}
m \\ k
\end{array} \right)
(m-k)^{m-1}=
\sum^{m}_{k=0}
(-1)^k 
\left( \begin{array}{c}
m \\ k
\end{array} \right)
(m-k)^{m-1}=S_3(m,m)=0.
\label{Am-1}
\end{equation}

This ends the first part of the proof.

\hspace{0.5 cm}2. Let's prove that if Eq.~(\ref{IntLTMPL}) is valid for     $n-m \le k<n$ ($n>m$) then, given Eq.~(\ref{Mstep}), it is also valid for $k=n$. Eq.~(\ref{Mstep}) can be written as
\begin{eqnarray}
&& x_n=\sum^{m-1}_{k=0}A(m-1,k)G_K(x_{n-k-1},h)- \sum^{m}_{k=1} (-1)^k 
\left( \begin{array}{c}
m \\ k
\end{array} \right) x_{n-k}=S_{1n}-S_{2n}.
\label{MstepNa}
\end{eqnarray}
Using the definition of $A(n,k)$, Eq.~(\ref{EN}), in $S_{1n}$ and substituting summation index $k$ by $j=n-k-1$ we have
\begin{eqnarray}
&& S_{1n}=\sum^{n-1}_{j=n-m} G_K(x_{j},h)\sum^{n-j-1}_{k=0} (-1)^k 
\left( \begin{array}{c}
m \\ k
\end{array} \right) (n-j-k)^{m-1}.
\label{MstepN1a}
\end{eqnarray}
Using Eq.~(\ref{IntLTMPL}) and changing the order of summation in $S_{2n}$ we have
\begin{eqnarray}
&& S_{2n}=\sum^{n-m-1}_{j=0} G_K(x_{j},h)\sum^{m}_{k=1} (-1)^k 
\left( \begin{array}{c}
m \\ k
\end{array} \right) (n-j-k)^{m-1} \nonumber \\
&&+\sum^{n-2}_{j=n-m} G_K(x_{j},h)\sum^{n-j-1}_{k=1} (-1)^k 
\left( \begin{array}{c}
m \\ k
\end{array} \right) (n-j-k)^{m-1}.
\label{MstepN2a}
\end{eqnarray}
Now Eq.~(\ref{MstepNa}) can be written as 
\begin{eqnarray}
&& x_{n}=\sum^{n-1}_{j=n-m} (n-j)^{m-1} G_K(x_{j},h)-
\sum^{n-m-1}_{j=0} G_K(x_{j},h)\sum^{m}_{k=1} (-1)^k 
\left( \begin{array}{c}
m \\ k
\end{array} \right) (n-j-k)^{m-1} \nonumber \\
&&=\sum^{n-1}_{j=0} (n-j)^{m-1} G_K(x_{j},h)- \sum^{n-m-1}_{j=0} G_K(x_{j},h)\sum^{m}_{k=0} (-1)^k 
\left( \begin{array}{c}
m \\ k
\end{array} \right) [(n-j)-k]^{m-1}.
\label{MstepN3a}
\end{eqnarray}
Using binomial formula and Eq.~(\ref{S4}) it is easy to prove that the last sum is equal zero.  

This ends the proof of Theorem~\ref{The1}.

\section{Maps with Real Power-Law Memory}
\label{Real}

Let's consider the following total usually used to define the Gr$\ddot{u}$nvald-Letnikov
fractional derivative (see \cite{Samko,Podlubny}):
\begin{eqnarray}
&&\sum^{n}_{k=0}
(-1)^k 
\left( \begin{array}{c}
\alpha \\ k
\end{array} \right)
x_{n-k}
= (-1)^n
\left( \begin{array}{c}
\alpha \\ n
\end{array} \right)
x_0 +
\sum^{n-1}_{k=0}
(-1)^k 
\left( \begin{array}{c}
\alpha \\ k
\end{array} \right)
\sum^{n-k-1}_{i=0}
(n-k-i)^{\alpha-1}G_K(x_i,h) 
= \nonumber \\ 
&&(-1)^n
\left( \begin{array}{c}
\alpha \\ n
\end{array} \right)
x_0 +
\sum^{n-1}_{i=0}G_K(x_i,h) 
\sum^{n-i-1}_{k=0}
(-1)^k 
\left( \begin{array}{c}
\alpha \\ k
\end{array} \right)
(n-k-i)^{\alpha-1}
=  \label{RealProof}
 \\ 
&&(-1)^n
\left( \begin{array}{c}
\alpha \\ n
\end{array} \right)
x_0 +
\sum^{n-1}_{i=0}G_K(x_i,h) A(\alpha-1,n-i-1), \nonumber
\end{eqnarray} 
where $\alpha$ is a real number. 
Transformation from the first to the second line in
Eq.~(\ref{RealProof}) requires changing of the order of summations
and can be seen on the same Fig.~{\ref{sum}} if one assumes $m=n-1$. 
We used the standard definition (see \cite{Samko,Podlubny})
\begin{equation}
\left( \begin{array}{c}
\alpha \\ n
\end{array} \right)
=
\frac{\alpha(\alpha-1)...(\alpha-n+1)}{n!}=\frac{\Gamma(\alpha+1)}
{\Gamma(n+1)\Gamma(\alpha-n+1)} 
\label{FracCombinations}
\end{equation} 
and the definition of the Eulerian numbers with fractional order parameters
introduced in \cite{BH}
\begin{equation}
A(\alpha,k)=
\sum^{k}_{j=0}
(-1)^j 
\left( \begin{array}{c}
\alpha+1 \\ j
\end{array} \right)
(k+1-j)^{\alpha}.
\label{FracEulerian}
\end{equation} 

Validity of Eq.~(\ref{LTMPL}) for $n=1$ follows from Eq.~(\ref{RealProof})
with $n=1$. If we assume that Eq.~(\ref{LTMPL}) is true for $k \le n$,
then from  Eq.~(\ref{RealProof}) written for $n+1$ follows
\begin{eqnarray}
&&x_{n+1}=
-\sum^{n}_{s=1}
(-1)^s 
\left( \begin{array}{c}
\alpha \\ s
\end{array} \right)
\sum^{n-s}_{k=0}
(n-s-k+1)^{\alpha-1}G_K(x_k,h) 
\nonumber \\ 
&&+
\sum^{n}_{k=0}G_K(x_k,h) 
\sum^{n-k}_{s=0}
(-1)^s
\left( \begin{array}{c}
\alpha \\ s
\end{array} \right)
(n-k-s+1)^{\alpha-1}
  \label{RealProofNN}
 \\ 
&&=
-\sum^{n-1}_{k=0}
G_K(x_k,h) 
\sum^{n-k}_{s=1}(-1)^s 
\left( \begin{array}{c}
\alpha \\ s
\end{array} \right)
(n-s-k+1)^{\alpha-1}           
\nonumber \\ 
&&+
\sum^{n}_{k=0}G_K(x_k,h) 
\sum^{n-k}_{s=0}
(-1)^s
\left( \begin{array}{c}
\alpha \\ s
\end{array} \right)
(n-s-k+1)^{\alpha-1}
= \sum^{n}_{k=0}(n-k+1)^{\alpha-1}G_K(x_k,h). \nonumber
\end{eqnarray}

Now we may formulate the following theorem:
\begin{theorem}
Any long term memory
map 
\begin{equation}
x_{n}=\sum^{n-1}_{k=0}(n-k)^{\alpha-1} G_K(x_k,h), \ \ (n>0)
\label{FrLTMPL}
\ee
where $\alpha \in  \mathbb{R}$ and  $n \in  \mathbb{N}$, 
is equivalent to the map 
\begin{equation}
\sum^{n}_{k=0}
(-1)^k 
\left( \begin{array}{c}
\alpha \\ k
\end{array} \right)
x_{n-k}=
(-1)^n
\left( \begin{array}{c}
\alpha \\ n
\end{array} \right)
x_0 +
\sum^{n-1}_{k=0}G_K(x_{n-k-1},h) A(\alpha-1,k).
\label{nstep}
\end{equation}
\label{The2}
\end{theorem}
For $n=0$  Eq.~(\ref{nstep}) yields the identity $x_0=x_0$ and 
for $n=1$ it yields $x_1=G_K(x_0,h)$  (notice that $A(\alpha,0)=1$).
In the case of a positive integer $\alpha=m$ Eq.~(\ref{nstep}) is equivalent to
(in the case $n>m$)  Eq.~(\ref{Mstep}). This follows from the the following:
\begin{equation}
\left( \begin{array}{c}
m \\ k
\end{array} \right)
=0 \ \ \ \mbox{for $(k>m)$, $A(m-1,k)=0$ for  $k > m-1$, and  Eq.~(\ref{Ajj})}.
\label{nstepIntAlpsLessn}
\end{equation}
The property $A(m-1,k)=0$ for  $k > m-1$ follows from Eq.~(\ref{Ajj}) 
and repeated applications of the recurrence formula Eq.~(\ref{ENRec}):
diagonal elements $A(j,j)$ are equal to zero and each element $A(n,k)$ is 
a linear combination of the elements to the left $A(n,k-1)$ and below
$A(n+1,k)$ with respect to this element.

\section{Behavior of Systems with Real Power-Law Memory}
\label{Beh}

\subsection{Discrete Systems}
\label{Disc}

For any finite $h$, systems with power-law memory are discrete
systems. Their behavior for $\alpha >0$ was preliminarily investigated in papers
\cite{DNC,Chaos,Chaos2014,DNC2014,ICFDA2014,ME1,ME2,ME3,ME4,ME5}. 
In the most important for biological
applications cases, $0<\alpha < 2$, the investigation is more detailed and is
done on the examples of the fractional Standard and Logistic maps. Maps
with  $m-1<\alpha \le m$, where $m \in  \mathbb{N}$, are equivalent to 
m-dimensional maps. For integer values of $\alpha=m>1$ these maps 
are m-dimensional volume preserving maps with no (one-step) memory.
It is easy to see that after the introduction 
 \begin{equation}
\begin{array}{c}
x^{(0)}_k=x_k, \\ x^{(1)}_k=x^{(0)}_k-x^{(0)}_{k-1}, \\ ...,\\
x^{(r)}_k=x^{(r-1)}_k-x^{(r-1)}_{k-1}, \\ ...,\\ x^{(m-1)}_k=x^{(m-2)}_k-x^{(m-2)}_{k-1},
\end{array} 
\label{mDimVar}
\end{equation}
where $k \ge m-1$, the map Eq.~(\ref{Mstep}) can be written as
\begin{equation}
\begin{array}{c}
\left\{
\begin{array}{lll}
x^{(m-1)}_n = x^{(m-1)}_{n-1}+
\sum^{m-2}_{k=0}A(m-1,k)
G_K\left(\sum^{k}_{i=0}(-1)^i
\left( \begin{array}{c}
k \\ i
\end{array} \right)
x^{(i)}_{n-1},h \right) \\ = x^{(m-1)}_{n-1}+F\left(x^{(0)}_{n-1},...,x^{(m-2)}_{n-1}\right),
\\ x^{(m-2)}_{n}=x^{(m-2)}_{n-1}+
x^{(m-1)}_{n},\\ ..., \\ 
x^{(m-k)}_{n}=x^{(m-k)}_{n-1}+x^{(m-k+1)}_{n},
\\ ...,\\x^{(0)}_{n}=x^{(0)}_{n-1}+x^{(1)}_{n}.
\end{array}
\right.
\end{array} 
\label{mDimEq}
\end{equation} 
The Jacobian  matrix ($m\times m$) of this
transformation 
$J_{(x_{n+1}^{(0)},x_{n+1}^{(1)},...,x_{n+1}^{(m-1)})}
(x_{n}^{(0)},x_{n}^{(1)},...,x_{n}^{(m-1)})$ is
\[ \left| \begin{array}{ccccccccc}
1+\frac{\partial F}{\partial x_n^{(0)}} & 1+\frac{\partial F}{\partial
  x_n^{(1)}} & 1+\frac{\partial F}{\partial x_n^{(2)}}
& ... & 1+\frac{\partial F}{\partial x_n^{(m-2)}} & 1  \\
\frac{\partial F}{\partial x_n^{(0)}} & 1+\frac{\partial F}{\partial
  x_n^{(1)}} & 1+\frac{\partial F}{\partial x_n^{(2)}}
& ... & 1+\frac{\partial F}{\partial x_n^{(m-2)}} & 1  \\
\frac{\partial F}{\partial x_n^{(0)}} & \frac{\partial F}{\partial
  x_n^{(1)}} & 1+\frac{\partial F}{\partial x_n^{(2)}}
& ... & 1+\frac{\partial F}{\partial x_n^{(m-2)}} & 1  \\
... & ... & ... & ... & ... &
...  \\
\frac{\partial F}{\partial x_n^{(0)}} & \frac{\partial F}{\partial
  x_n^{(1)}} & \frac{\partial F}{\partial x_n^{(2)}}
& ... & 1+\frac{\partial F}{\partial x_n^{(m-2)}} & 1  \\
\frac{\partial F}{\partial x_n^{(0)}} & \frac{\partial F}{\partial
  x_n^{(1)}} & \frac{\partial F}{\partial x_n^{(2)}}
& ... & \frac{\partial F}{\partial x_n^{(m-2)}} & 1  \end{array} \right| . \]
The first column of this matrix can be written as the sum of 
the column with one in the
first row and the remaining zeros and the column which is equal to
${\partial F}/{\partial x_n^{(0)}}$ times the last column. The determinant
of the latter one is zero. It is easy to show
recursively that determinant of the former one is equal to one and the map 
Eq~(\ref{mDimEq}) indeed is volume preserving.

As it has been shown in paper  \cite{DNC},
the complexity of the behavior of discrete systems with  positive 
power law memory increases with the increase in power. When the power is 
fractional, systems demonstrate the new types of behavior which include the new 
types of attractors and the non-uniqueness of solutions. 
The new types of attractors
include cascade of bifurcations types trajectories (CBTT) and intermittent
CBTT. As a result of the non-uniqueness, attractors may overlap and  
phase space trajectories intersect. Systems with $\alpha \le 0$ are not 
investigated.

\subsection{Continuous Systems}
\label{Cont}

Let's assume, according to the general approach in the definition  of
the  Gr$\ddot{u}$nvald-Letnikov fractional derivative, that  
\begin{equation}
x=x(t), \ \  x_k=x(t_k), \ \ t_k=a+kh, \ \ nh=t-a 
\label{GLDef}
\end{equation}
for $0 \le k \le n$. 
 If one divides Eq.~(\ref{Mstep}) by $h^m$ in the case of positive integer
values of $\alpha$ and considers a limit $h \rightarrow 0+$, then the left
side of the resulting equation will give the $m^{th}$ derivative from $x(t)$ 
at the time $t$. If we assume 
\begin{equation}
G_K(x,h)=\frac{1}{\Gamma (\alpha)}h^{\alpha}G_K(x), 
\label{FORxSmooth}
\end{equation}
where $G_K(x)$ is continuous, then $x(t) \in C^m$. 
The map Eq.~(\ref{LTMPL}) can be written as
\begin{equation}
x(t)=\frac{1}{\Gamma (\alpha)} h\sum^{n-1}_{k=0,nh=t-a}(t-t_k)^{\alpha-1} G_K(x(t_k)) 
\label{LTMPLh}
\end{equation}
and in the limit $h \rightarrow 0$  Theorem~\ref{The1} can be
formulated as a well-known result
\begin{theorem}
The Volterra integral equation of the second kind
\begin{equation}
x(t)=\frac{1}{\Gamma (m)}
\int^{t}_{a}\frac{G_K(x(\tau))d\tau}{(t-\tau)^{1-m}}, \  \ \ (t>a) 
\label{VoltLTMPLh}
\end{equation}
where $m \in  \mathbb{N}$ and $G_K(x) \in C^0$ on the range 
$D \in  \mathbb{R}$ of the 
function $x(t)$ ($t \in [a,b]$), is equivalent on $[a,b]$ 
to the differential equation
\begin{equation}
\frac{d^m
x(t)}{dt^m}=\frac{1}{\Gamma (m)}\sum^{m-1}_{k=0}A(m-1,k)G_K(x(t))=G_K(x(t)),
\label{MstepC}
\end{equation}
where we used the classical result $\sum^{m-2}_{k=0}A(m-1,k)=\Gamma(m)$,
with the zero initial conditions 
\begin{equation}
c_k=\frac{d^kx(t)}{dt^k}(t=a)=0, \  \  \ k=0,1,...,m-1. 
\label{IC}
\end{equation}
\label{The3}
\end{theorem}

While discrete equations Eqs.~(\ref{IntLTMPL})~and~(\ref{Mstep}) have a unique
solutions for any function $G_K(x)$, the corresponding continuous equations
Eqs.~(\ref{VoltLTMPLh})~and~(\ref{MstepC}) 
require the Lipschitz condition on $G_K(x)$ in $D$. Because this is not
essential for this paper, in what follows we always assume that the 
$G_K(x)$ satisfies the Lipschitz condition  in $D$.

In the case $c_k \ne 0$ the well-known equivalence of the differential equation Eq.~(\ref{MstepC}) to the  Volterra integral equation of the second kind
\begin{equation}
x(t)= \sum^{m-1}_{k=0}\frac{c_k}{\Gamma(k+1)}(t-a)^k+\frac{1}{\Gamma (m)}
\int^{t}_{a}\frac{G_K(x(\tau))d\tau}{(t-\tau)^{1-m}}, \  \ \ (t>a) 
\label{VoltNonZero}
\ee
follows in the limit $h \rightarrow 0$  from the generalization of Theorem~\ref{The1}:
\begin{theorem}
Any long term memory
map 
\begin{equation}
x_{n}=\sum^{m-1}_{k=0}\frac{c_k}{\Gamma(k+1)}(nh)^{k} +
\frac{h^{m}}{\Gamma(m)}\sum^{n-1}_{k=0}(n-k)^{m-1} G_K(x_k), \ \ (n>0)
\label{IntLTMPLN}
\end{equation}
where $m \in  \mathbb{N}$, is equivalent to the $m$-step memory map 
\begin{eqnarray}
&& x_{n}=\sum^{m-1}_{k=0}\frac{c_k}{\Gamma(k+1)}(nh)^{k} +
\frac{h^{m}}{\Gamma(m)}\sum^{n-1}_{k=0}(n-k)^{m-1} G_K(x_k), \ \ (0<n \le m),
\nonumber   \\
&&\sum^{m}_{k=0} (-1)^k 
\left( \begin{array}{c}
m \\ k
\end{array} \right)
x_{n-k}= \frac{h^{m}}{\Gamma(m)}\sum^{m-1}_{k=0}A(m-1,k)G_K(x_{n-k-1})
, \ \ (n > m)
\label{MstepNW}
\end{eqnarray}
\label{The4}
\end{theorem}
\proof
The proof of this theorem is similar to the proof of Theorem~\ref{The1}.

1. The first part of the proof uses the fact that for $n>m$
$m^{th}$ backward difference of the first sum in  Eq.~(\ref{IntLTMPLN})
is equal to zero:
\begin{equation}
\sum^{m}_{k=0} (-1)^k 
\left( \begin{array}{c}
m \\ k
\end{array} \right)
\sum^{m-1}_{i=0}\frac{c_i}{\Gamma(i+1)}[(n-k)h]^{i} =
\sum^{m-1}_{i=0}\frac{c_ih^i}{\Gamma(i+1)}
\sum^{m}_{k=0} (-1)^k 
\left( \begin{array}{c}
m \\ k
\end{array} \right)
(n-k)^{i}.
\label{MstepN2}
\end{equation}
After we apply the binomial formula to $(n-k)^i$ and use the identity
Eq.~(\ref{S4}) it is clear that the internal sum on the right hand side
(RHS) is equal to zero. 

2. In the second part of the proof an additional term on the RHS
of Eq.~(\ref{MstepN2a}) is
\begin{eqnarray}
&&\sum^{m}_{k=1} (-1)^k 
\left( \begin{array}{c}
m \\ k
\end{array} \right)
\sum^{m-1}_{i=0}\frac{c_i}{\Gamma(i+1)}[(n-k)h]^{i} =
\sum^{m-1}_{i=0}\frac{c_ih^i}{\Gamma(i+1)}
\sum^{m}_{k=1} (-1)^k 
\left( \begin{array}{c}
m \\ k
\end{array} \right)
(n-k)^{i} \nonumber   \\
&&=-\sum^{m-1}_{i=0}\frac{c_i}{\Gamma(i+1)}(nh)^i,
\label{MstepN3}
\end{eqnarray}
which completes the proof of Theorem~\ref{The4}.

3. From Eq.~(\ref{MstepNW}) follows that $x(a)=x_0=c_0$ and for $0<n<m$
\begin{eqnarray}
&&x^{(n)}(a)=\lim_{h \rightarrow 0}\frac{1}{h^n}\sum^{n}_{k=0} (-1)^k 
\left( \begin{array}{c}
n \\ k
\end{array} \right) x_{n-k}=
\lim_{h \rightarrow 0}\frac{1}{h^n}\sum^{n}_{k=0} (-1)^k 
\left( \begin{array}{c}
n \\ k
\end{array} \right)
\sum^{m-1}_{i=0}\frac{c_ih^i}{\Gamma(i+1)}(n-k)^i \nonumber   \\
&&=\lim_{h \rightarrow 0}\frac{1}{h^n}\sum^{m-1}_{i=0}\frac{c_ih^i}
{\Gamma(i+1)}
\sum^{n}_{k=0} (-1)^k 
\left( \begin{array}{c}
n \\ k
\end{array} \right)
(n-k)^{i}=c_n.
\label{ICNN}
\end{eqnarray}
In the last sum all terms with $i<n$ are zeros because of Eq.~(\ref{S4});
limit $h \rightarrow 0$ of all terms with $i>n$ is also zero;
when $i=n$ the only term which gives non-zero sum over $k$ in the binomial
expansion of $(n-k)^n$ is $(-1)^nk^n$ and the corresponding sum is $n!$.

As we mentioned in Sec.~\ref{int}, a transition from discrete 
to continuous dynamical system in the case 
$m=2$ results in the disappearance of chaos, which, in general, should not
be the case for systems with non-degenerate memory and for the case, which
is important 
in applications, $0<\alpha<2$, we may expect that corresponding 
continuous systems will still have chaotic solutions. 

Let's consider the limit $h \rightarrow 0$ for fractional 
$\alpha>0$ in Eq.~(\ref{nstep}) divided by $h^{\alpha}$ given  Eq.~(\ref{GLDef})
\begin{equation}
\lim_
{\begin{array}{c}
n \rightarrow \infty \\ nh=t-a
\end{array}}
h^{-\alpha}
\left\{
\sum^{n}_{k=0}
(-1)^k 
\left( \begin{array}{c}
\alpha \\ k
\end{array} \right)
x_{n-k}=
(-1)^n
\left( \begin{array}{c}
\alpha \\ n
\end{array} \right)
x_0 +\frac{1}{\Gamma(\alpha)}
\sum^{n-1}_{k=0}h^{\alpha}G_K(x_{n-k-1}) A(\alpha-1,k)
\right\}.
\label{nstepLim}
\end{equation}
The LHS of Eq.~(\ref{nstepLim}) coincides with the definition of the 
Gr$\ddot{u}$nvald-Letnikov fractional derivative:
\begin{equation}
\lim_
{\begin{array}{c}
n \rightarrow \infty \\ nh=t-a
\end{array}}
h^{-\alpha}\sum^{n}_{k=0}
(-1)^k 
\left( \begin{array}{c}
\alpha \\ k
\end{array} \right)
x_{n-k}=
\lim_
{\begin{array}{c}
h \rightarrow 0 \\ nh=t-a
\end{array}}
h^{-\alpha}\sum^{n}_{k=0}
(-1)^k 
\left( \begin{array}{c}
\alpha \\ k
\end{array} \right)
x(t-kh)=
_aD^{\alpha}_tx(t),
\label{ContLeft}
\end{equation}
where $x(t)$ is assumed to be $\lceil \alpha \rceil $ times continuously differentiable on $[a,t]$.
The first term on the RHS of Eq.~(\ref{nstepLim}) is equal to zero:
\begin{equation}
\lim_
{
\begin{array}{c}
n \rightarrow \infty \\ nh=t-a
\end{array} 
}
h^{-\alpha}
(-1)^n
\left( \begin{array}{c}
\alpha \\ n
\end{array} \right)
x_{0}=(-1)^nx_0
(t-a)^{-\alpha}
\lim_{n \rightarrow \infty}
n^{\alpha}
\left( \begin{array}{c}
\alpha \\ n
\end{array} \right)
\label{ContRight1}
\end{equation}
and 
\begin{eqnarray}
&&\lim_{n \rightarrow \infty}
\left| n^{\alpha}
\left( \begin{array}{c}
\alpha \\ n
\end{array} \right)
\right|=
\lim_{n \rightarrow \infty}
\left|
\frac{n^{\alpha}\Gamma(\alpha+1)}{n!\Gamma(1-(n-\alpha))}
\right|
=\left|\frac{\Gamma(\alpha+1)\sin(\pi \alpha)}{\pi}\right|
\lim_{n \rightarrow \infty}
\frac{n^{\alpha}\Gamma(n-\alpha)}{n!}
=\nonumber \\ 
&&\left|\frac{\Gamma(\alpha+1)\sin(\pi \alpha)}{\pi}\right|
\lim_{n \rightarrow \infty}
\frac{n^{\alpha}n^{-\alpha}(n-1)!}{n!}
=
\left|\frac{\Gamma(\alpha+1)\sin(\pi \alpha)}{\pi}\right|
\lim_{n \rightarrow \infty}\frac{1}{n}
=0.
\label{ContRight1Add}
\end{eqnarray}
Here we used the well known properties of the Gamma-function:
$\Gamma(1-z)\Gamma(z)=\pi / \sin(\pi z)$ and 
$\lim_{n \rightarrow \infty}\Gamma(n+\alpha)/[\Gamma(n)n^{\alpha}]=1$.

The evaluation of the last term in Eq.~(\ref{nstepLim}) will 
require some revision of the results obtained in  \cite{BH,Westphal}:
\begin{enumerate}
\item The
last theorem (Theorem 9) proven in \cite{BH}, which states that for any
$\alpha>1$ and $k \in \mathbb{N}_0$
\begin{equation}
A(\alpha,k)=\Gamma(\alpha +1)\int^{k+1}_{k}p_{\alpha}(x)dx,
\label{BH9}
\end{equation}
\begin{equation}
\sum^{\infty}_{k=0}A(\alpha,k)=\Gamma(\alpha+1),
\label{ASum}
\end{equation}
where
\begin{equation}
p_\alpha(x):=
\left\{ 
\begin{array}{c}
0, \\ \frac{1}{\Gamma(\alpha)}
\sum^{}_{0 \le j < x}
(-1)^j
\left( \begin{array}{c}
\alpha \\ j
\end{array} \right)
(x-j)^{\alpha-1},
\end{array}\right.
\begin{array}{c}
-\infty < x \le 0 \\ 0<x<\infty 
\end{array}
\label{pAlp}
\end{equation}
is based on the results from  \cite{Westphal} which are obtained for
$\alpha>0$. The one line proof of Theorem 9 in  \cite{BH} is nowhere 
violated for $0< \alpha \le 1$. Thus, we assume that
Eqs.~(\ref{BH9})~and~(\ref{ASum}) are true for $\alpha>0$. 
\item 
According to the asymptotic formula for large $k$ 
from the fifth page of \cite{Westphal}
for $\alpha>0$, integer $k$, and $0 < \Theta \le 1$
\begin{equation}
p_{\alpha}(k+\Theta)=O(k^{-\alpha-1})\Theta^{\alpha-1}+O(k^{-\alpha-1}+k^{\alpha-[\alpha]-2}).
\label{BH9n}
\end{equation}
Then 
\begin{equation}
A(\alpha-1,k)=
\sum^{k}_{j=0}
(-1)^j 
\left( \begin{array}{c}
\alpha \\ j
\end{array} \right)
(k+1-j)^{\alpha-1}=\Gamma(\alpha)p_{\alpha}(k+1)=O(k^{-\alpha-1}+k^{\alpha-[\alpha]-2}).
\label{AP}
\end{equation} 
As a continuous function $x(\tau)$ attains its maximum 
$x_{max}$ and minimum  $x_{min}$ values on $[a,t]$ 
and is bounded ($|x|<M_1$). Assuming that $G_K(x)$ is a 
continuous function on  $[x_{min},x_{max}]$, this 
function is also bounded ($|G_K(x)|<M_2$). This yields
\begin{equation}
\lim_
{n \rightarrow \infty} 
\sum^{n-1}_{k=0}|G_K(x_{n-k-1}) A(\alpha-1,k)| \le
\lim_
{n \rightarrow \infty} 
\sum^{n-1}_{k=0}M_2O(k^{-\alpha-1}+k^{\alpha-[\alpha]-2}) 
<\infty.
\label{LastTermConverge}
\end{equation}
\end{enumerate}
Now, for $\alpha>0$ we may write 
\begin{eqnarray}
&&\lim_
{\begin{array}{c}
n \rightarrow \infty \\ nh=t-a
\end{array}}
\sum^{n}_{k=0}G_K(x_{n-k})A(\alpha-1,k) \nonumber \\
&&=
\lim_
{\begin{array}{c}
n \rightarrow \infty \\ nh=t-a
\end{array}}
\sum^{N_1}_{k=0}G_K(x(t-\frac{k}{n}(t-a)))A(\alpha-1,k)+
\sum^{\infty}_{k=N_1+1}G_K(x_{n-k})A(\alpha-1,k),
\label{ThirdTerm}
\end{eqnarray}
where 
for an
arbitrarily small $\varepsilon>0$ there exists $N$ such that for 
$\forall$ $N_1>N$ the
following holds
\begin{equation}
\left| \sum^{\infty}_{k=N_1+1}G_K(x_{n-k})A(\alpha-1,k)  \right| 
<  \frac{\varepsilon}{2}. 
\label{ALimit}
\end{equation}
In Eq.~(\ref{ThirdTerm}) by choosing $n>N_2>>N_1$ the argument 
of the function $x(\tau)$ in the first sum  
on the right can be made arbitrarily close to $t$ so that due to the
continuity of $x(\tau)$ and $G_K(x)$
\begin{equation}
\sum^{N_1}_{k=0}\left[ G_K(x(t-\frac{k}{n}(t-a)))-G_K(x(t) \right] 
A(\alpha-1,k)< \frac{\varepsilon}{2}.
\label{Gdif}
\end{equation}
Eqs.~(\ref{LastTermConverge})-(\ref{Gdif}) yield
\begin{equation}
\lim_
{\begin{array}{c}
n \rightarrow \infty \\ nh=t-a
\end{array}}
\sum^{n}_{k=0}G_K(x_{n-k})A(\alpha-1,k)=
G_K(x(t)\lim_{n \rightarrow \infty}
\sum^{n}_{k=0}A(\alpha-1,k),
\label{ThirdTermNN}
\end{equation}
where the series on the right converges absolutely for $\alpha>0$ 
according to  Eq.~(\ref{AP}). 
According to Eqs.~(\ref{ASum})~and~(\ref{MstepC}) for $\alpha \ge 1$ 
the sum on the right is equal to $\Gamma(\alpha)$ and in the limit $h \rightarrow \infty$ we may formulate the
following theorem: 

\begin{theorem}
For $\alpha \in  \mathbb{R}$, $\alpha \ge 1$ 
The Volterra integral equation of the second kind
\begin{equation}
x(t)=\frac{1}{\Gamma (\alpha)}
\int^{t}_{a}\frac{G_K(x(\tau))d\tau}{(t-\tau)^{1-\alpha}}, \  \ \ (t>a) 
\label{VoltReal}
\ee
where $G_K(x(\tau))$ is a continuous on $x \in [x_{min}(\tau), x_{max}(\tau)]$,
$\tau\in[a,t]$ function is equivalent to the fractional differential equation 
\begin{equation}
_aD^{\alpha}_tx(t)=G_K(x(t)),
\label{Theorem3b}
\end{equation}
where the derivative on the left is the Gr$\ddot{u}$nvald-Letnikov
fractional derivative,
with the zero initial conditions 
\begin{equation}
c_k=\frac{d^kx(t)}{dt^k}(t=a)=0, \  \  \ k=0,1,...,\lceil \alpha \rceil-1. 
\label{IC1}
\end{equation}
\label{The5}
\end{theorem}

The methods used 
in  \cite{BH,Westphal} do not allow us to prove Eq.~(\ref{ASum})   
for $-1<\alpha<0$
but based on the convergence of the series in Eq.~(\ref{ThirdTermNN}) 
we'll formulate the following conjecture:

\begin{conj} 
Theorem~\ref{The5} is valid for $0<\alpha<1$. 
\label{Con6}
\end{conj}

Theorem~\ref{The5} and Conjecture~\ref{Con6} is not a new result. 
It is known (see \cite{Samko,Podlubny,KilBook}) that Riemann-Liouville and 
Caputo derivatives coincide in the case $c_k={d^kx(t)}/{dt}(t=a)=0$,  
$k=0,1,..., [\alpha] $ and also that for $x(t) \in C^{[\alpha]}[a,T]$ and 
integrable $x^{[\alpha]+1}(t)$ in $[a,T]$ ($a<t<T$) Riemann-Liouville 
and Gr$\ddot{u}$nvald-Letnikov fractional derivatives 
$_aD^{\alpha}_tx(t)$ coincide. 

For $t>a$ the left-sided Riemann-Liouville fractional derivative is defined as
\be
_a^{RL}D^{\alpha}_t x(t)=D^n_t \ _aI^{n-\alpha}_t x(t)=
\frac{1}{\Gamma(n-\alpha)} \frac{d^n}{dt^n} \int^{t}_a 
\frac{x(\tau) d \tau}{(t-\tau)^{\alpha-n+1}},
\label{RL}
\ee
where $n-1 \le \alpha < n$, $\alpha \in  \mathbb{R}$, 
$n \in  \mathbb{N}$, $D^n_t=d^n/dt^n$, and $ _0I^{\alpha}_t$ 
is a Riemann-Liouville fractional integral.
In the definition of the left-sided Caputo fractional derivative the order
of integration and differentiation is switched: 
\be
_a^CD^{\alpha}_t x(t)=\ _aI^{n-\alpha}_t \ D^n_t x(t) 
=\frac{1}{\Gamma(n-\alpha)}  \int^{t}_a 
\frac{ D^n_{\tau}x(\tau) d \tau}{(t-\tau)^{\alpha-n+1}}.
\label{Cap}
\ee
In \cite{KM1,KM2} Kilbas and Marzan showed that fractional 
differential equation 
\begin{equation}
_a^CD^{\alpha}_tx(t)=G_K(t,x(t)), \  \ 0 <\alpha, \  \  t\in[a,T]
\label{KMCap}
\end{equation}
with the initial conditions 
\begin{equation}
\frac{d^kx(t)}{dt^k}(t=a)=c_k, \  \  \ k=0,1,...,\lceil \alpha \rceil-1 
\label{IC2}
\end{equation}
is equivalent to the Volterra integral equation of the second kind 
\begin{equation}
x(t)= \sum^{\lceil \alpha \rceil-1 }_{k=0}\frac{c_k}{\Gamma(k+1)}(t-a)^k+\frac{1}{\Gamma (\alpha)}
\int^{t}_{a}\frac{G_K(\tau,x(\tau))d\tau}{(t-\tau)^{1-\alpha}}, \  \ \ (t>a) 
\label{VoltRealN}
\end{equation}
in the space $C^{\lceil \alpha \rceil-1 }[a,T]$. 
A similar result for
the equivalence of the equation with the Riemann-Liouville 
fractional derivative 
\begin{equation}
_a^{RL}D^{\alpha}_tx(t)=G_k(t,x(t)), \  \ 0 <\alpha
\label{KMRL}
\end{equation}
with the initial conditions 
\begin{equation}
(_a^{RL}D^{\alpha-k}_tx) (a+)=c_k, \  \  \ k=1,2,...,\lceil \alpha \rceil
\label{IC3}
\end{equation} 
to the Volterra integral equation of the second kind 
\begin{equation}
x(t)= \sum^{\lceil \alpha \rceil}_{k=1}\frac{c_k}{\Gamma(\alpha-k+1)}(t-a)^{\alpha-k}+\frac{1}{\Gamma (\alpha)}
\int^{t}_{a}\frac{G_K(\tau,x(\tau))d\tau}{(t-\tau)^{1-\alpha}}, \  \ \ (t>a) 
\label{VoltRealNN}
\end{equation}
for $x(t) \in  L(a,T)$ and $G(t,x(t)) \in  L(a,T)$ was proved by Kilbas, Bonilla, and Trujillo in \cite{KBT1,KBT2}.

On one hand, in the case of $x(t) \in C^{\lceil \alpha \rceil-1 }[a,T]$
and the zero initial conditions all above defined derivatives are equivalent
and Eq.~(\ref{Theorem3b}) is equivalent to Eq.~(\ref{VoltReal}). On the other
hand we saw that for $\alpha>0$  Eq.~(\ref{VoltReal}) is equivalent 
(see Eq.~(\ref{ThirdTermNN})) to 
\begin{equation}
_aD^{\alpha}_tx(t)=
\frac{1}{\Gamma(\alpha)}G_K(x(t)\lim_{n \rightarrow \infty}
\sum^{n}_{k=0}A(\alpha-1,k).
\label{ThirdTermNNN}
\end{equation}   
This proves Conjecture~\ref{Con6} and Eq.~(\ref{ASum}) for $\alpha>-1$. 

We'll end this section with the theorem which in the limit $h \rightarrow
0$ yields the equivalence of problem Eq.~(\ref{KMRL})~and~Eq.~(\ref{IC3})
to the problem Eq.~(\ref{VoltRealNN}) in the case 
$c_{\lceil \alpha \rceil}=0$, which corresponds to a finite value of $x(a)$:

\begin{theorem}
Any long term memory map 
\begin{equation}
x_{n}=\sum^{\lceil \alpha \rceil - 1}_{k=1}\frac{c_k}{\Gamma(\alpha-k+1)}
(nh)^{\alpha-k}  +\sum^{n-1}_{k=0}(n-k)^{\alpha-1} G_K(x_k,h),
\label{FrLTMPLNN}
\ee
where $\alpha \in  \mathbb{R}$, is equivalent to the map 
\begin{eqnarray}
&&\sum^{n}_{k=0}
(-1)^k 
\left( \begin{array}{c}
\alpha \\ k
\end{array} \right)
x_{n-k}-\sum^{\lceil \alpha \rceil - 1}_{i=1}\frac{c_ih^{\alpha-i}}
{\Gamma(\alpha-i+1)}
\sum^{i-1}_{k=0}
(-1)^k 
\left( \begin{array}{c}
i-1 \\ k
\end{array} \right)
A(\alpha-i,n-k-1) \nonumber \\
&&=(-1)^n
\left( \begin{array}{c}
\alpha \\ n
\end{array} \right)
x_0 +
\sum^{n-1}_{k=0}G_K(x_{n-k-1},h) A(\alpha-1,k).
\label{nstepNN}
\end{eqnarray}
\label{The7}
\end{theorem}
\proof
1. The first part of the proof is the same as the proof of
Theorem~\ref{The2} plus the following result:
\begin{eqnarray}
&&\sum^{n}_{k=0}
(-1)^k 
\left( \begin{array}{c}
\alpha \\ k
\end{array} \right)
\sum^{\lceil \alpha \rceil - 1}_{i=1}\frac{c_i}{\Gamma(\alpha-i+1)}
[(n-k)h]^{\alpha-i}=
\sum^{\lceil \alpha \rceil - 1}_{i=1}\frac{c_ih^{\alpha-i}}{\Gamma(\alpha-i+1)}
\sum^{n-1}_{k=0}
(-1)^k 
\left( \begin{array}{c}
\alpha \\ k
\end{array} \right)
(n-k)^{\alpha-i} \nonumber \\
&&=\sum^{\lceil \alpha \rceil - 1}_{i=1}\frac{c_ih^{\alpha-i}}
{\Gamma(\alpha-i+1)}
\sum^{i-1}_{k=0}
(-1)^k 
\left( \begin{array}{c}
i-1 \\ k
\end{array} \right)
A(\alpha-i,n-k-1).
\label{Long1}
\end{eqnarray}
Here we used the identity  
\begin{eqnarray}
&&\sum^{n-1}_{k=0}
(-1)^k 
\left( \begin{array}{c}
\alpha \\ k
\end{array} \right)
(n-k)^{\alpha-i}=
\sum^{n-1}_{k=0}
(-1)^k 
\sum^{i-1}_{j=0}
\left( \begin{array}{c}
i-1 \\ j
\end{array} \right)
\left( \begin{array}{c}
\alpha -i +1 \\ k-j
\end{array} \right)
(n-k)^{\alpha - i} \nonumber \\
&&=\sum^{i-1}_{j=0}
\left( \begin{array}{c}
i-1 \\ j
\end{array} \right)
\sum^{n-1}_{k=j}
(-1)^k 
\left( \begin{array}{c}
\alpha -i +1 \\ k-j
\end{array} \right)
(n-k)^{\alpha - i} \nonumber \\ 
&&=\sum^{i-1}_{j=0}(-1)^j 
\left( \begin{array}{c}
i-1 \\ j
\end{array} \right)
\sum^{n-j-1}_{k=0}
(-1)^k 
\left( \begin{array}{c}
\alpha -i +1 \\ k
\end{array} \right)
(n-k-j)^{\alpha - i} 
\nonumber \\
&&=\sum^{i-1}_{k=0}
(-1)^k 
\left( \begin{array}{c}
i-1 \\ k
\end{array} \right)
A(\alpha-i,n-k-1), \   \  0<i<\lceil \alpha \rceil.
\label{Short1}
\end{eqnarray}

2. Eq.~(\ref{nstepNN}) with $n=1$ yields  Eq.~(\ref{FrLTMPLNN}).
 If we assume that Eq.~(\ref{FrLTMPLNN}) is true for $k \le n$,
then we may write the equation for $x_{n+1}$ as in  Eq.~(\ref{RealProofNN})
with two additional terms on the RHS:

\begin{eqnarray}
&&x_{n+1}=\sum^{n}_{k=0}(n-k+1)^{\alpha-1}G_K(x_k,h)+
\sum^{\lceil \alpha \rceil - 1}_{i=1}\frac{c_ih^{\alpha-i}}
{\Gamma(\alpha-i+1)}
\sum^{i-1}_{k=0}
(-1)^k 
\left( \begin{array}{c}
i-1 \\ k
\end{array} \right)
A(\alpha-i,n-k) \nonumber \\
&&-\sum^{n}_{k=1}
(-1)^k 
\left( \begin{array}{c}
\alpha \\ k
\end{array} \right)
\sum^{\lceil \alpha \rceil - 1}_{i=1}\frac{c_ih^{\alpha-i}}
{\Gamma(\alpha-i+1)}(n+1-k)^{\alpha-i}=
\sum^{n}_{k=0}(n-k+1)^{\alpha-1}G_K(x_k,h) \nonumber \\
&&+\sum^{\lceil \alpha \rceil - 1}_{i=1}\frac{c_ih^{\alpha-i}}
{\Gamma(\alpha-i+1)}
\sum^{i-1}_{k=0}
(-1)^k 
\left( \begin{array}{c}
i-1 \\ k
\end{array} \right)
A(\alpha-i,n-k)  \label{Last} \\
&&-\sum^{\lceil \alpha \rceil - 1}_{i=1}\frac{c_ih^{\alpha-i}}
{\Gamma(\alpha-i+1)}
 \Bigl[
\sum^{n}_{k=0}
(-1)^k 
\left( \begin{array}{c}
\alpha \\ k
\end{array} \right)
(n+1-k)^{\alpha-i}-(n+1)^{\alpha-i}\Bigr] \nonumber \\
&&=\sum^{\lceil \alpha \rceil - 1}_{k=1}\frac{c_k}{\Gamma(\alpha-k+1)}
[(n+1)h]^{\alpha-k}  +\sum^{n}_{k=0}(n-k+1)^{\alpha-1} G_K(x_k,h). \nonumber 
\end{eqnarray}

3. From fractional calculus it is known that the Gr$\ddot{u}$nvald-Letnikov 
 fractional derivative of the power function $f(t)=(t-a)^{\beta}$ is
\begin{equation}
_aD^{\alpha}_t (t-a)^{\beta}=
\lim_
{\begin{array}{c}
n \rightarrow \infty \\ nh=t-a
\end{array}}
h^{-\alpha}\sum^{n}_{k=0}(-1)^k 
\left( \begin{array}{c}
\alpha \\ k
\end{array} \right)[(n-k)h]^{\beta}=
\frac{\Gamma(\beta+1)}{\Gamma(-\alpha+\beta+1)}(t-a)^{\beta-\alpha},
\label{Last1}
\end{equation}
where $\alpha<0$, $\beta>-1$ or $0 \le m \le \alpha < m+1$, $\beta>m$  
(see Sec. 2.2.4 in \cite{Podlubny}). This yields for $\beta=\alpha-i$, 
$i \in \mathbb{Z}$, and
$\beta, \alpha>0$
\begin{equation}
\lim_
{\begin{array}{c}
n \rightarrow \infty \\ nh=t-a
\end{array}}
h^{-\alpha}\sum^{n}_{k=0}(-1)^k 
\left( \begin{array}{c}
\alpha \\ k
\end{array} \right)
[(n-k)h]^{\beta}=
\left\{ \begin{array}{c}
{\Gamma(\beta+1)}(t-a)^{-i}/(-i)!, \ \ i<0; \\ 
\Gamma(\beta+1),   \ \ \ \  \  \  \ \   \ i=\alpha-\beta=0; \\
0, \  \ \ \ \ \    \ \   \ i>0.
\end{array} \right.
\label{Last2}
\end{equation}
For $k=1,2,...,{\lceil \alpha \rceil - 1}$ Eq.~(\ref{FrLTMPLNN}) leads to 
\begin{eqnarray}
&&_aD^{\alpha}_t x(a+)=\lim_{t \rightarrow a+} \lim_
{\begin{array}{c}
n \rightarrow \infty \\ nh=t-a
\end{array}} 
h^{k-\alpha}\sum^{n}_{j=0}(-1)^j 
\left( \begin{array}{c}
\alpha -k \\ j
\end{array} \right)
x_{n-j}  \nonumber \\
&&=\lim_{t \rightarrow a+} \lim_
{\begin{array}{c}
n \rightarrow \infty \\ nh=t-a
\end{array}} 
h^{k-\alpha}\sum^{n}_{j=0}(-1)^j 
\left( \begin{array}{c}
\alpha -k \\ j
\end{array} \right)
\sum^{\lceil \alpha \rceil - 1}_{i=1}\frac{c_i}{\Gamma(\alpha-i+1)}
[(n-j)h)^{\alpha-i} \nonumber \\
&&=\sum^{\lceil \alpha \rceil - 1}_{i=1}\frac{c_i}{\Gamma(\alpha-i+1)}
\lim_{t \rightarrow a+} \lim_
{\begin{array}{c}
n \rightarrow \infty \\ nh=t-a
\end{array}} 
h^{k-\alpha}\sum^{n}_{j=0}(-1)^j 
\left( \begin{array}{c}
\alpha -k \\ j
\end{array} \right)
[(n-j)h)^{\alpha-i} \nonumber \\
&&=\sum^{\lceil \alpha \rceil - 1}_{i=1}\frac{c_i}{\Gamma(\alpha-i+1)}
\left\{ \begin{array}{c}
{\lim_{t \rightarrow a+}\Gamma(\alpha-i+1)}(t-a)^{k-i}/(k-i)!, \ \ k>i; \\ 
\Gamma(\alpha-i+1),   \ \ \ \  \  \  \ \   \ i=k; \\
0, \  \ \ \ \ \    \ \   \ k<i.
\end{array} \right. 
= c_k  
\label{Last3}
\end{eqnarray}

The direct calculation of the LHS of Eq~(\ref{Last2}) with $m=-i \ge
0$ yields
\begin{eqnarray}
&&\lim_
{\begin{array}{c}
n \rightarrow \infty \\ nh=t-a
\end{array}}
h^{-\alpha}\sum^{n}_{k=0}(-1)^k 
\left( \begin{array}{c}
\alpha \\ k
\end{array} \right)[(n-k)h]^{\beta}=
\lim_
{\begin{array}{c}
n \rightarrow \infty \\ nh=t-a
\end{array}}
h^{m}\sum^{n}_{k=0}(-1)^k 
\left( \begin{array}{c}
\beta-m \\ k
\end{array} \right)(n-k)^{\beta} \nonumber \\
&&=\lim_
{\begin{array}{c}
n \rightarrow \infty \\ nh=t-a
\end{array}}
h^{m}\sum^{n}_{k=0}(-1)^k (n-k)^{\beta}
\sum^{k}_{j_0=0}(-1)^{j_0}
\left( \begin{array}{c}
\beta-m+1 \\ k-j_0
\end{array} \right) \nonumber \\
&&=(t-a)^m\lim_{n \rightarrow \infty}n^{-m} 
\sum^{n-1}_{j_0=0}\sum^{n-j_0-1}_{k=0}(-1)^k
\left( \begin{array}{c}
\beta-m+1 \\ k
\end{array} \right) 
(n-j_0-k)^{\beta} \nonumber \\
&&=(t-a)^m\lim_{n \rightarrow \infty}n^{-m} 
\sum^{n-1}_{j_0=0}\sum^{j_0}_{k=0}(-1)^k
\left( \begin{array}{c}
\beta-m+1 \\ k
\end{array} \right) 
(j_0+1-k)^{\beta} \nonumber \\
&&=(t-a)^m\lim_{n \rightarrow \infty}n^{-m} \sum^{n-1}_{j_0=0}
\sum^{j_0}_{j_1=0}\sum^{j_1}_{j_2=0}...
\sum^{j_m}_{k=0}(-1)^k
\left( \begin{array}{c}
\beta+1 \\ k
\end{array} \right) 
(j_m+1-k)^{\beta} \nonumber \\
&&=(t-a)^m\lim_{n \rightarrow \infty}n^{-m} \sum^{n-1}_{j_0=0}
\sum^{j_0}_{j_1=0}\sum^{j_1}_{j_2=0}...
\sum^{j_{m-1}}_{j_m=0}A(\beta,j_m)
=\frac{1}{m!}(t-a)^m\lim_{n \rightarrow \infty}
\sum^{n-1}_{s=0}\frac{\Gamma(m+n-s)}{n^m\Gamma(n-s)}A(\beta,s)\nonumber \\
&&=\frac{1}{m!}
(t-a)^m\lim_{n \rightarrow \infty}\sum^{n-1}_{s=0}D(m,n,s)A(\beta,s)
=\frac{1}{m!}
(t-a)^m\lim_{n \rightarrow \infty}S_n
=\frac{1}{m!}\Gamma(\beta+1)(t-a)^{m}. 
\label{Last4}
\end{eqnarray}
The transition within the sixth line of this chain of transformations
is based on the Theorem 1 from \cite{DNC2014}, which states that for 
$\forall n \in \mathbb{N}$     
\begin{equation} 
_a\Delta^{-n}_{t}f(t)=\frac{1}{(n-1)!} \sum^{t-n}_{s=a}(t-s-1)^{(n-1)}
f(s)=\sum^{t-n}_{s^0=a} \sum^{s^0}_{s^1=a}...
\sum^{s^{n-2}}_{s^{n-1}=a}f(s^{n-1}),
\label{MRInt}
\end{equation}
where falling factorial function $t^{(\alpha)}$ is defined as
\begin{equation}
t^{(\alpha)} =\frac{\Gamma(t+1)}{\Gamma(t+1-\alpha)}.
\label{FrFac}
\end{equation}
For $m=0$ the  equality 
\begin{equation}
\lim_{n \rightarrow \infty}
\sum^{n-1}_{s=0}\frac{\Gamma(m+n-s)}{n^m\Gamma(n-s)}A(\beta,s)=
\Gamma(\beta+1)
\label{Last5}
\end{equation} 
coincides with Eq.~(\ref{ASum}), which is true for $\beta>-1$. 
Series $\sum^{n-1}_{s=0}A(\beta,s)$ converges absolutely and  
$D(m,n,s)$, which is a product of m factors
\begin{equation}
D(m,n,s)= (1-\frac{s}{n})(1-\frac{s-1}{n})...(1-\frac{s-m+1}{n}) <
(1+\frac{m}{n})^m,
\label{Last6}
\end{equation} 
is bounded. This means that $S_n$ converges absolutely to some $S$. 
For  $\forall \varepsilon >0 $ there  exists $N_1$ such that for
$\forall N \ge N_1$ simultaneously
$|\sum^{N_2-1}_{s=N_1}D(m,N_2,s)A(\beta,s)|<\varepsilon/3$ and $|\Gamma(\beta+1)-\sum^{N_1-1}_{s=0}A(\beta,s)|<\varepsilon/3$. For $N_2 >>N_1$ and $s \le N_1$ 
\begin{equation}
1-m\frac{N_1}{N_2}<(1-\frac{N_1}{N_2})^m<D(m,N_2,s)<(1+\frac{N_1}{N_2})^m
< 1+\frac{m^2}{N_2}+o(\frac{m^2}{N_2})  
\label{Last7}
\end{equation}
and 
\begin{equation}
|D(m,N_2,s)-1|<m\frac{N_1}{N_2}.
\label{Last8}
\end{equation}
For $\forall N_2 > N_{\varepsilon}$, where 
\begin{equation}
N_{\varepsilon}=\frac{3mN_1\sum^{\infty}_{s=0}|A(\beta,s)|}{\varepsilon},
\label{Last9}
\end{equation}
we can write
\begin{eqnarray}
&&|S_{N_2}-\Gamma(\beta+1)|=
\Bigl|\sum^{N_2-1}_{s=0}D(m,N_2,s)A(\beta,s)-\Gamma(\beta+1)\Bigr|
< 
\Bigl| \sum^{N_2-1}_{s=N_1}D(m,N_2,s)A(\beta,s)| \nonumber \\
&&+\sum^{N_1-1}_{s=0}|D(m,N_2,s)-1||A(\beta,s)|+
\Bigl|\sum^{N_1-1}_{s=0}A(\beta,s)-\Gamma(\beta+1)\Bigr| < \varepsilon.
\label{Last10}
\end{eqnarray}
This means that $S=\Gamma(\beta+1)$.

If in Eq.~(\ref{Last2}) $i>0$, then using Eq.~(\ref{Short1}), we may write
\begin{eqnarray}
&&\lim_
{\begin{array}{c}
n \rightarrow \infty \\ nh=t-a
\end{array}}
h^{-\alpha}\sum^{n}_{k=0}(-1)^k 
\left( \begin{array}{c}
\alpha \\ k
\end{array} \right)[(n-k)h]^{\beta}=
\lim_
{\begin{array}{c}
n \rightarrow \infty \\ nh=t-a
\end{array}}
h^{-i}\sum^{n}_{k=0}(-1)^k 
\left( \begin{array}{c}
\alpha \\ k
\end{array} \right)(n-k)^{\alpha-i} \nonumber \\
&&=(t-a)^{-i}\lim_{n \rightarrow \infty}n^i
\sum^{i-1}_{k=0}
(-1)^k 
\left( \begin{array}{c}
i-1 \\ k
\end{array} \right)
A(\alpha-i,n-k-1).
\label{Last11}
\end{eqnarray}
Comparing  Eq.~(\ref{Last11}) to  Eq.~(\ref{Last2}) we may formulate a new
property of Eulerian numbers:
\begin{equation}
\lim_{n \rightarrow \infty}n^i
\sum^{i-1}_{k=0}
(-1)^k 
\left( \begin{array}{c}
i-1 \\ k
\end{array} \right)
A(\alpha-i,n-k-1)=0, \ \ (i>0). 
\label{Last12}
\end{equation}

\section{Summary}
\label{Summary}

Here we summarize the main results obtained in this paper. We start with  
the fractional difference calculus. Theorem~\ref{The2} can be formulated as the
equivalence of  maps with power-law memory (power $\alpha-1$) 
generated by a function $G_K(x,h)$, where $x$ is the map's variable, $K$ 
is a parameter, and $h$ is the map's step 
(constant time between two consecutive iterations),  
to fractional difference equations in which 
Gr$\ddot{u}$nvald-Letnikov like fractional difference operator acting on
the map's variable on the LHS is equal to the convolution of the values 
of the generating function from all previous steps $k$ with the Eulerian
numbers $A(\alpha-1,k)$ on the RHS. In the case of the integer power-law
memory this theorem can be formulated as a simpler result
(Theorem~\ref{The1}): any long term non-negative
integer power-law memory (power $m-1$) 
map is equivalent to a m-step memory map
(the $m^{th}$ backward difference on the LHS is equal to the convolution of 
the generating functions from the $MAX(1,m-1)$ previous values of the map's
variable  with the Eulerian numbers $A(m-1,k)$ on the RHS).
Maps with long term posititve integer ($m>1$) power-law memory are equivalent
to m-dimensional volume preserving maps with no (one-step) memory.

In the continuous limit ($h \rightarrow 0$) 
Theorems~\ref{The1}~and~\ref{The2}
yield the well-known results of the equivalence of 
differential equations to the
integral Volterra equations of the second kind in both integer 
and fractional cases. In the process of transition to the continuous limit
we were able to prove that the property of Eulerian numbers 
$\sum^{\infty}_{k=0}A(\alpha,k)=\Gamma(\alpha+1)$, Eq.(\ref{ASum}),
known for  $ \alpha > 1$,  is true for $ \alpha > -1$ and obtained a new
property of Eulerian numbers Eq.~(\ref{Last12}).

\section{Conclusion}
\label{Conclusion}

Phase space of discrete  non-linear integer maps with power-law memory
may demonstrate islands of stability and chaotic areas.
These maps are well investigated for $m=2$ but 
investigation of general properties of such maps for $m>2$ is far from
completion. 
Eq.~(\ref{alp1}) yields the regular logistic map if we assume
$G_K(x,h)=-G_K^L(x)=-x+Kx(1-x)$. Eq.~(\ref{mDimEq}) with 
$G_K(x,h)=-G_K^{SM}(x)=-K\sin(x)$ yields the regular standard map.
This is why we'll call maps 
Eqs.~(\ref{IntLTMPL}),~(\ref{Mstep}),~(\ref{FrLTMPLNN}),~and~(\ref{nstepNN})  
with $G_K(x,h)=-G_K^{L}(x)$ the logistic maps with memory or the fractional 
logistic maps and with  
$G_K(x,h)=-G_K^{SM}(x)$ the standard maps with memory or the fractional 
standard maps.
Initial investigation of maps with long term 
fractional  power-law memory in 
\cite{DNC,Chaos,Chaos2014,DNC2014,ICFDA2014,ME1,ME2,ME3,ME4,ME5} 
has been done on the examples of the fractional 
logistic and standard maps with $0< \alpha <3$. 
New types of attractors (CBTT) were obtained for $0< \alpha <2$.  

If we consider 
Eq.~(\ref{nstepNN}) with $G_K(x,h)=h^{\alpha}KG(x)$, then,
up to the term depending on the initial conditions,
solution of this fractional difference  equation depends only on the 
product  $h^{\alpha}K$. 
This type of systems includes fractional standard map ($G(x)=-\sin(x)$)
and a system, which in the limit $h \rightarrow 0$ yields the fractional
logistic differential equation ($G(x)=x(1-x)$).
In the case $h=1$ for $0<\alpha<2$ the 
fractional standard and logistic maps with  
$|K| \lesssim 1$ have only sinks (see Fig.~\ref{Fig1}a) (no chaos). 
We may conclude that for small $h$ there will be no chaotic trajectories
for $|K| \lesssim h^{-\alpha}$, which implies a  possibility 
that in the limit  $h \rightarrow 0$ the fractional logistic differential
equation and the limit of the fractional standard map 
($D^{\alpha}x(t)/Dt^{\alpha}=K\sin(x)$) will have no chaotic solutions
for $0 < \alpha< 2$. This kind of reasoning may not work for all
fractional systems.
The stability of the $x=1$ fixed point of the
fractional logistic differential equation also follows from the elementary
stability analysis (see, e.g., \cite{LDE1}). 
In  \cite{CFDE}, on the basis of the analysis of two fractional order 
autonomous non-linear systems, authors conjectured that chaos may exist  
in  autonomous non-linear systems with a total system's order of 
$2+\varepsilon$, where $0<\varepsilon<1$. Examples of fractional chaotic
attractors in continuous systems of the order less than three can be found
also  in \cite{ZSE}. 

To the best of our knowledge, there is no proof that chaos can't exist
in fractional systems of the order less than two.
To prove it or to find a counterexample is a challenging problem.
Another challenging problem is to investigate if there are analogs of  
cascade of bifurcations type trajectories in continuous systems.

\section*{Acknowledgments}
The author expresses his gratitude to Eliezer Hameiri  and the
administration of the Courant Institute of Mathematical Sciences
for the opportunity to complete this work at Courant,  
to Harold Weitzner and Vasily Tarasov for useful remarks, and to Virginia  Donnelly
for technical help.

\end{document}